**Schrödinger's Equation as a Consequence of the Central Limit Theorem without Prior Assumption of Physical Laws**

P. M. Grinwald
Melbourne, Australia
Email: paulgrinwald@gmail.com

The central limit theorem has been found to apply to random vectors in complex Hilbert space. This amounts to sufficient reason to study the complex–valued Gaussian, looking for relevance to quantum mechanics. Here we show that the Gaussian, with all terms fully complex, acting as a propagator, leads to Schrödinger's nonrelativistic equation including scalar and vector potentials, assuming only that the norm is conserved. No physical laws need to be postulated *a priori*. It thereby presents as a process of irregular motion analogous to the real random walk but executed under the rules of the complex number system. Inferences are 1: There is a standard view that Schrödinger's equation is deterministic, while wavefunction "collapse" is probabilistic (by Born's rule) -- this is opposed by the now-demonstrated linkage to the central limit theorem, indicating a stochastic picture for the foundation of Schrödinger's equation itself. 2: This picture is also consistent with the dynamic origin of probabilities suggested for the Born rule in the de Broglie-Bohm pilot-wave theory. Reasons for the primary role of **C** are open to discussion. The present derivation is compared with recent reconstructions of the quantum formalism, which have the aim of rationalizing its obscurities.

Key words: Time-dependent Schrödinger equation; Feynman action formula; Path integrals; Gaussian processes; Central limit theorem; Random walks; Pilot-wave theory

# 1. Introduction

We are often reminded (e.g. Ref. [1]) that the Schrödinger equation was postulated empirically, being justified only *post hoc* by its agreement with quantum mechanical observations. This equation might appear less strange or arbitrary if it could be reasoned from something more basic.

We take the Gaussian density function acting as a propagator, but instead of using it in the real number system **R**, we let it act in the complex number system **C**. Why **C**? **C** has a more complete logic [2] (see Discussion, **3.1**). This is such a general quality that in appealing to it, consistency would require us to make all terms complex, not just some arbitrary selection of variables. We do so. We will note reasons for preferring **C** rather than other division algebras (**3.1**).

The initial motivation for this model was to build a better understanding of Feynman's path integral approach [3, 4]. One feels that Feynman's action formula, though mysterious in itself, must contain essential physical insight since it leads to Schrödinger's equation, while his use of Gaussian integrals seems just a mathematical tool. Yet the Gaussian seems a good place to start, because the real Gaussian is familiar and well understood. So before going to the action formula, it is reasonable to prepare the ground by looking at (a) the Gaussian in **R**, then (b) the Gaussian in **C**, then (c) the complex Gaussian as a propagator, and then (d) the complex Gaussian propagator that is norm-conserving. One might guess that (d) could be a fairly complicated object. How close does (d) get us to Schrödinger's equation, before we even introduce Feynman's action formula?

The surprising result is that the action formula is not needed at all. After developing the Gaussian as in (d) above, there will be no call on any information from physics (de Broglie relations; Newton's laws, force fields, energy conservation, or classical action with or without Feynman's arbitrary formulation; not even Galilean kinematics). We reach Schrödinger's equation (nonrelativistic with scalar and vector potentials).

In effect we use the intuitively appealing path integral method with a more general starting point than usual. Further, we point out that it is essential for consistency to calculate the normalization constant correct to first order in the time increment, not merely zero order as in Feynman [3, 4]. The need for any input of physical information is thereby removed.

Since the derivation depends only on the Gaussian, the physical content of Feynman's action formula, and of the Schrödinger equation itself, is seen to reside in the Gaussian when the latter is closely considered. Taking this a step further, our focus on the Gaussian is given a strong rationale by the fact that the central limit theorem is known to apply to complex Hilbert space vectors (see Discussion, **3.1**); the central limit theorem of course leads to Gaussians.

Thus Schrödinger's equation and the physical laws that can be derived from it are mere reflections of laws concerning the behaviour of pure numbers within the complex number system. That is the central result of this article.

Derivation of these results is given in **2**. In **3**, the Discussion, we consider how to interpret them. The real Gaussian is well known to be associated with a random walk in many manifestations. What does it mean if the Gaussian is complex instead of real? In **3.1** we suggest it may still be understood as a random process, but one that is executed according to the rules of **C** instead of **R**. In **3.2** we add that this picture is consistent with the usual textbook teaching of quantum mechanics [1], including Born's statistical interpretation (and is unlike the type of random walk described by Nelson and others). In **3.3** we compare with previous justifications (Schrödinger [5], Feynman [3, 4], Kac [6], Nelson [7], Jauch [8] and others), and with modern reconstructions of quantum mechanics (**3.3.1**). Conclusions are summarised in **4**.

# 2. Derivation

A preliminary overview of the subsections is as follows:

2.1 Real Gaussian: We recall the form of the real Gaussian and its normalization constant.

2.2 A "convenient" complex Gaussian: We want all the terms in the Gaussian to be complex in general, with the Gaussian acting to propagate another function $\psi$. However the general case gives rise to many possibilities that cannot be dealt with in a single concise manner. We therefore undertake a multi-part procedure. As a first stage we will arrive at the essential equations, including normalization, using the most "convenient" case for the composition of the Gaussian. This part of the derivation follows the path integral method [3, 4]. The other cases will be dealt with later (see **2.4**).

2.3 Correction to first order in the time increment: Feynman [3] and Feynman and Hibbs [4] calculated the normalization constant correct to *zero* order in the small time increment $\varepsilon$, but in their working they take Taylor expansions up to *first* order in $\varepsilon$. This seems inconsistent. To ensure consistency therefore, we make the normalization constant correct to first order in $\varepsilon$, so that it matches the level of approximations in the Taylor expansion.

2.4 Check alternatives: With equations in hand for the "convenient" case, we check all the alternative cases for normalizability.

2.5 Schrödinger equation: Finally we write the admissible (i.e. normalizable) equations for the propagator and the corresponding differential equation. The latter takes the form of the Schrödinger equation. No assumptions based on physical observations will have been made in the course of the derivation. (We have used symbols such as $x$ and $t$ for variables to foreshadow their later physical interpretation, but that is merely a convenient notation and does not imply any prior laws of physics relating to the variables.)

We now set this out in detail.

## 2.1 Real Gaussian

The general form of the Gaussian distribution for one variable is

$$P(x) = \frac{1}{K}\exp-\frac{(x-\mu)^2}{2\sigma^2}, \tag{1}$$

where $P$ is the density at the point $x$, the mean and variance are $\mu$ and $\sigma^2$ respectively, and $K$ is the normalization constant [9].

An equation in this form can represent a time-dependent process,

$$P(x,t) = \frac{1}{K}\exp-\frac{(x-ut)^2}{2Dt}, \tag{2}$$

where $t$ is the time, $u$ is the drift velocity, and $D$ is the diffusion constant. The mean is now $ut$ and the variance is $Dt$. Putting $\int P\,dx = 1$, where the integral is taken over all space, we have

$$K = [2\pi Dt]^{1/2}. \tag{3}$$

A Gaussian in the form of equation (2) may be written as a propagator [10],

$$\Pi(\eta,\varepsilon;x,t) = \frac{1}{K}\exp-\frac{(\eta-u(x,t)\varepsilon)^2}{2D(x,t)\varepsilon}, \tag{4}$$

where at a point $(x,t)$, $\Pi$ dictates the step length $\eta$ in the small time interval $\varepsilon$. The drift $u$ and the diffusivity $D$ are allowed to vary with $x$ and $t$. Acting at each point on a density function $P(x,t)$, the propagator determines how the local density will change with time. $K$ retains the same form as above, now written

$$K = [2\pi D(x,t)\varepsilon]^{1/2}. \tag{5}$$

## 2.2 A "convenient" complex Gaussian

We intend to see what happens when all the parameters and variables in the Gaussian propagator are allowed to be complex. This makes for a complicated picture. It is convenient to start with one particular case, and to look at the alternatives later.

Specifically, we replace the real diffusivity $D$ by the pure imaginary diffusivity $iD$ (with $D$ real). Also we keep $D$ constant (no dependence on $x$ or $t$). In this way equation (4) is replaced by

$$\Pi(\eta,\varepsilon;x,t) = \frac{1}{K}\exp\frac{i\left(\eta - u\left(x,t\right)\varepsilon\right)^2}{2D\varepsilon}. \tag{6}$$

We remark that equation (4) has been justified using random variable theory [10], which applies only to real systems because of its axioms. In looking at the complex counterpart, equation (6), we recognize that the same justification does not apply, but at this stage we are merely exploring an interesting equation. Justification in its own right will be discussed in **3.1**.

Equation (6) is related to the propagator in Feynman's path integral approach [3, 4]. Taking up that approach, we calculate $\psi(x,t+\varepsilon)$ as the sum of contributions transferred from values of $\psi$ situated nearby at a slightly earlier time, that is, from $\psi(x+\eta,t)$. Those transfers are calculated from the propagator $\Pi(\eta, \varepsilon; x+\eta, t)$, hence

$$\psi\left(x,t+\varepsilon\right) = \int \Pi(\eta,\varepsilon;x+\eta,t)\psi(x+\eta,t)d\eta. \tag{7}$$

From equations (6) and (7),

$$\psi\left(x,t+\varepsilon\right) = \int \frac{1}{K}\exp\frac{i\left(\eta - u\left(x+\eta,t\right)\varepsilon\right)^2}{2D\varepsilon}\psi(x+\eta,t)d\eta. \tag{8}$$

To abbreviate some notation in equation (8), let

$$u = u(x,t), \tag{9}$$

$$u_+ = u(x+\eta,t). \tag{10}$$

Using this notation in equation (8), the exponential term is given by

$$\exp \frac{i(\eta - u_+ \varepsilon)^2}{2D\varepsilon} = \exp \frac{i}{2D\varepsilon}\left(\eta^2 - 2\eta u_+ \varepsilon + u_+^2 \varepsilon^2\right)$$

$$= \exp \frac{i\eta^2}{2D\varepsilon} \exp - \frac{i\eta u_+}{D} \exp \frac{iu_+^2 \varepsilon}{2D}. \qquad (11)$$

When $\eta$ and $\varepsilon$ approach zero, we have $\eta u_+ / D \to 0$ and $u_+^2 \varepsilon / 2D \to 0$, given that $u_+$ and $D$ are finite (non-zero). Hence we may write Taylor expansions of the second and third exponentials on RHS in equation (11),

$$\exp - \frac{i\eta u_+}{D} = 1 - \frac{i\eta u_+}{D} - \frac{\eta^2 {u_+}^2}{2D^2}, \qquad (12)$$

$$\exp \frac{i{u_+}^2 \varepsilon}{2D} = 1 + \frac{i{u_+}^2 \varepsilon}{2D}, \qquad (13)$$

keeping terms only to second order in $\eta$ and first order in $\varepsilon$ (as we will do throughout, following Feynman [3] and Feynman and Hibbs [4]). From equations (11), (12) and (13),

$$\exp \frac{i(\eta - u_+ \varepsilon)^2}{2D\varepsilon} = \left[1 - \frac{i\eta u_+}{D} - \frac{\eta^2 {u_+}^2}{2D^2} + \frac{i{u_+}^2 \varepsilon}{2D}\right] \exp \frac{i\eta^2}{2D\varepsilon}. \qquad (14)$$

These authors [3, 4] point out that a term such as the exponential containing $\eta^2/\varepsilon$ on RHS of equation (8) oscillates rapidly with small $\varepsilon$ except near $\eta$=0, and the rapid oscillations would contribute little to the integration on $\eta$ due to cancellations. Since appreciable contributions to the integral are then expected only for small $\eta$, Taylor expansion of $\psi(x + \eta, t)$ is justified. Substituting equation (14) in equation (8) and making Taylor expansions of $\psi(x, t + \varepsilon)$ and $\psi(x + \eta, t)$,

$$\psi(x,t) + \varepsilon \frac{\partial \psi}{\partial t} = \int d\eta \frac{1}{K}\left[1 - \frac{i\eta u_+}{D} - \frac{\eta^2 u_+^2}{2D^2} + \frac{iu_+^2 \varepsilon}{2D}\right]\left(\exp \frac{i\eta^2}{2D\varepsilon}\right) \times \left[\psi(x,t) + \eta \frac{\partial \psi}{\partial x} + \frac{1}{2}\eta^2 \frac{\partial^2 \psi}{\partial x^2}\right]$$

$$(15)$$

Recalling equations (9), (10), with Taylor expansion for small $\eta$,

$$u_+ = u(x + \eta, t) = u(x,t) + \eta \frac{\partial u(x,t)}{\partial x} + \ldots = u + \eta \frac{\partial u}{\partial x} + \ldots. \qquad (16)$$

(Second order term is not shown as it will go to higher order when multiplied later.)

Using the integrals

$$\int_{-\infty}^{\infty} \exp\frac{i\eta^2}{2D\varepsilon} d\eta = (2\pi i D\varepsilon)^{1/2}, \tag{17}$$

$$\int_{-\infty}^{\infty} \eta \exp\frac{i\eta^2}{2D\varepsilon} d\eta = 0, \tag{18}$$

$$\int_{-\infty}^{\infty} \eta^2 \exp\frac{i\eta^2}{2D\varepsilon} d\eta = (2\pi i D\varepsilon)^{1/2} iD\varepsilon, \tag{19}$$

$$\int_{-\infty}^{\infty} \eta^4 \exp\frac{i\eta^2}{2D\varepsilon} d\eta = (2\pi i D\varepsilon)^{1/2} (-3) D^2\varepsilon^2, \tag{20}$$

we find that the two terms in equation (15) that involve $u_+^2$, when expanded by equation (16), cancel each other (to first order in $\varepsilon$) upon integration,

$$\int d\eta \left[ -\frac{\eta^2 u_+^{\ 2}}{2D^2} + \frac{iu_+^{\ 2}\varepsilon}{2D} \right] \left( \exp\frac{i\eta^2}{2D\varepsilon} \right) = \int d\eta u_+^{\ 2} \left[ -\frac{\eta^2}{2D^2} + \frac{i\varepsilon}{2D} \right] \left( \exp\frac{i\eta^2}{2D\varepsilon} \right)$$

$$= \int d\eta \left( u + \eta\frac{\partial u}{\partial x} \right)^2 \left[ -\frac{\eta^2}{2D^2} + \frac{i\varepsilon}{2D} \right] \left( \exp\frac{i\eta^2}{2D\varepsilon} \right)$$

$$= 0. \tag{21}$$

The only other term in equation (15) that involves $u_+$ is $-i\eta u_+ / D$. Expanding again with equation (16),

$$-\frac{i\eta u_+}{D} = -\frac{i\eta u}{D} - \frac{i\eta^2}{D}\frac{\partial u}{\partial x}. \tag{22}$$

With equation (21) and equation (22), equation (15) becomes

$$\psi(x,t) + \varepsilon\frac{\partial\psi}{\partial t} = \int d\eta \frac{1}{K} \left[ 1 - \frac{i\eta u}{D} - \frac{i\eta^2}{D}\frac{\partial u}{\partial x} \right] \left( \exp\frac{i\eta^2}{2D\varepsilon} \right) \times \left[ \psi(x,t) + \eta\frac{\partial\psi}{\partial x} + \frac{1}{2}\eta^2\frac{\partial^2\psi}{\partial x^2} \right]$$

$$= \int d\eta \psi \frac{1}{K} \left[ 1 - \frac{i\eta u}{D} - \frac{i\eta^2}{D}\frac{\partial u}{\partial x} \right] \exp\frac{i\eta^2}{2D\varepsilon}$$

$$+ \frac{\partial\psi}{\partial x}\frac{1}{K} \left[ \eta - \frac{i\eta^2 u}{D} \right] \exp\frac{i\eta^2}{2D\varepsilon}$$

$$+\frac{\partial^2\psi}{\partial x^2}\frac{1}{K}\left[\frac{1}{2}\eta^2\right]\exp\frac{i\eta^2}{2D\varepsilon}. \tag{23}$$

To evaluate the normalization factor *K*, we compare the leading $\psi(x,t)$ terms on the two sides. On the left-hand side there is simply $\psi(x,t)$. On the right-hand side, $\psi(x,t)$ is multiplied by the expression on the LHS of the following equation, which, from equation (17), is evaluated as

$$\frac{1}{K}\int_{-\infty}^{\infty}\exp\frac{i\eta^2}{2D\varepsilon}d\eta=\frac{1}{K}\left(2\pi iD\varepsilon\right)^{1/2}. \tag{24}$$

In order that both sides of equation (23) agree in the limit as $\varepsilon$ approaches zero, *K* must be chosen so that the expression in equation (24) equals 1; that is,

$$K=\left(2\pi iD\varepsilon\right)^{1/2}. \tag{25}$$

The equation is now correct to zero order in $\varepsilon$. This is the normalization factor given in the classic path integral formulation [3, 4] using the foregoing justification.

But since we have been taking Taylor expansions to first order in $\varepsilon$, (as in Refs. 3 and 4), we should not be satisfied to have the normalization constant specified only to zero order in $\varepsilon$. Consistency requires that we develop it to first order in $\varepsilon$.

## 2.3 Correction to first order in time increment

To do this, we insert into the normalization constant a first-order term in $\varepsilon$, making it as general as possible. So we amend equation (25) to

$$K=\left(2\pi iD\varepsilon\right)^{1/2}\left(1+\varepsilon T(x,t)\right), \tag{26}$$

where *T* is independent of $\psi$ (otherwise the propagator is not Gaussian), but may be some complex function of *x* and *t*, and $\varepsilon$ is small. The choice of sign for *T* is arbitrary.

Resuming in the manner of the Feynman exposition [3, 4], we use equation (26) and substitute

$$\frac{1}{K}=\left(2\pi iD\varepsilon\right)^{-1/2}\left(1-\varepsilon T(x,t)\right) \tag{27}$$

into equation (6) for the propagator, which becomes

$$\Pi(\eta,\varepsilon;x,t) = \left(2\pi iD\varepsilon\right)^{-1/2}\left(1-\varepsilon T(x,t)\right)\exp\frac{i(\eta - u(x,t)\varepsilon)^2}{2D\varepsilon}. \tag{28}$$

To develop the differential equation, we substitute equation (27) into equation (23). Included is a term involving $u$ which will later yield the vector potential $A$. Feynman warned that caution must be used with the integral $\int A dx$, possibly due to the individual trajectory of a single particle being undifferentiable ("like Brownian motion"), and introduced the midpoint rule which allows the integral to correspond with the known physics [3]. However, a single particle trajectory is unlike a density function, and we treat the latter as well behaved (Riemann integrable).

We proceed then to the integrations in equation (23), using equation (27) for $1/K$, and take from equation (19) that the $\eta^2$ terms in the integrand produce $iD\varepsilon$ terms in the integral,

$$\varepsilon\frac{\partial\psi}{\partial t} = \frac{\left(iD\varepsilon\right)}{2}\frac{\partial^2\psi}{\partial x^2} - \left(iD\varepsilon\right)\frac{iu}{D}\frac{\partial\psi}{\partial x} - \left(iD\varepsilon\right)\frac{i}{D}\frac{\partial u}{\partial x}\psi - \varepsilon T\psi. \tag{29}$$

Dividing through by $\varepsilon$, we get the differential equation

$$\frac{\partial\psi}{\partial t} = \frac{iD}{2}\frac{\partial^2\psi}{\partial x^2} + u\frac{\partial\psi}{\partial x} + \frac{\partial u}{\partial x}\psi - T\psi, \tag{30}$$

and its complex conjugate

$$\frac{\partial\psi^*}{\partial t} = -\frac{iD}{2}\frac{\partial^2\psi^*}{\partial x^2} + u\frac{\partial\psi^*}{\partial x} + \frac{\partial u}{\partial x}\psi^* - T^*\psi^*. \tag{31}$$

We already have normalization at time zero. We also require that equations (30, 31) conserve the norm over time,

$$\frac{d}{dt}\int_{-\infty}^{\infty}\left(\psi^*\psi\right)dx = 0. \tag{32}$$

To examine this, we expand the expression

$$\frac{d}{dt}\int_{-\infty}^{\infty}\left(\psi^*\psi\right)dx = \int_{-\infty}^{\infty}\frac{\partial}{\partial t}\left(\psi^*\psi\right)dx = \int_{-\infty}^{\infty}\left(\psi^*\frac{\partial\psi}{\partial t} + \frac{\partial\psi^*}{\partial t}\psi\right)dx$$

$$= \int_{-\infty}^{\infty} \psi^* \left( \frac{iD}{2} \frac{\partial^2 \psi}{\partial x^2} + u \frac{\partial \psi}{\partial x} + \frac{\partial u}{\partial x} \psi - T\psi \right) dx + \int_{-\infty}^{\infty} \left( -\frac{iD}{2} \frac{\partial^2 \psi^*}{\partial x^2} + \frac{\partial \psi^*}{\partial x} u + \psi^* \frac{\partial u}{\partial x} - \psi^* T^* \right) \psi dx$$

$$= \int_{-\infty}^{\infty} \left\{ \psi^* \frac{iD}{2} \frac{\partial^2 \psi}{\partial x^2} - \frac{iD}{2} \frac{\partial^2 \psi^*}{\partial x^2} \psi \right\} + \left\{ \psi^* u \frac{\partial \psi}{\partial x} + 2\psi^* \frac{\partial u}{\partial x} \psi + \frac{\partial \psi^*}{\partial x} u \psi \right\} + \left\{ -\psi^* T \psi - \psi^* T^* \psi \right\} dx$$

(33)

where we used equations (30, 31) in the second-last line, and rearranged terms for the last line.

Taking the first curly bracket on RHS of equation (33),

$$\int_{-\infty}^{\infty} \left\{ \psi^* \frac{iD}{2} \frac{\partial^2 \psi}{\partial x^2} - \frac{iD}{2} \frac{\partial^2 \psi^*}{\partial x^2} \psi \right\} dx = \frac{iD}{2} \int_{-\infty}^{\infty} \frac{\partial}{\partial x} \left\{ \psi^* \frac{\partial \psi}{\partial x} - \frac{\partial \psi^*}{\partial x} \psi \right\} dx$$

$$= \frac{iD}{2} \left[ \psi^* \frac{\partial \psi}{\partial x} - \frac{\partial \psi^*}{\partial x} \psi \right]_{-\infty}^{\infty}$$

$$= 0, \qquad (34)$$

the definite integral being zero with square-integrable $\psi$.

The second curly bracket on RHS contains the derivative of the triple product, which we separate,

$$\int_{-\infty}^{\infty} \psi^* u \frac{\partial \psi}{\partial x} + \psi^* \frac{\partial u}{\partial x} \psi + \frac{\partial \psi^*}{\partial x} u \psi dx + \int_{-\infty}^{\infty} \psi^* \frac{\partial u}{\partial x} \psi dx$$

$$= \int_{-\infty}^{\infty} \frac{\partial}{\partial x} \left( \psi^* u \psi \right) dx + \int_{-\infty}^{\infty} \psi^* \frac{\partial u}{\partial x} \psi dx$$

$$= \left[ \psi^* u \psi \right]_{-\infty}^{\infty} + \int_{-\infty}^{\infty} \psi^* \frac{\partial u}{\partial x} \psi dx$$

$$= \int_{-\infty}^{\infty} \psi^* \frac{\partial u}{\partial x} \psi dx, \qquad (35)$$

the quantity $\left[ \psi^* u \psi \right]_{-\infty}^{\infty} = 0$, again because $\psi$ is square-integrable.

For the last two terms on RHS, we write the real and imaginary parts of $T$ separately, $T = a + ib$, $T^* = a - ib$, with $a$ and $b$ real. Then

$$\int_{-\infty}^{\infty} -\left( \psi^* T \psi + \psi^* T^* \psi \right) dx = -\int_{-\infty}^{\infty} \psi^* \left( a + ib + a - ib \right) \psi dx$$

$$= -\int_{-\infty}^{\infty} \psi^* (2a) \psi dx. \tag{36}$$

Since equations (34), (35) and (36) must satisfy equation (32), we must have

$$\int_{-\infty}^{\infty} \psi^* \frac{\partial u}{\partial x} \psi dx - \int_{-\infty}^{\infty} \psi^* (2a) \psi dx = 0, \tag{37}$$

and hence

$$a(x,t) = \frac{1}{2} \frac{\partial u(x,t)}{\partial x}, \tag{38}$$

for all $x,t$. There is no restriction on $b(x,t)$ because it always cancels.

Thus $T$ is not necessarily zero. Re$T$ is zero only if $u$ is constant. And $T$ is determined only up to a phase factor.

To correct the normalization constant to first order in $\varepsilon$, we substitute

$$T(x,t) = \frac{1}{2} \frac{\partial u(x,t)}{\partial x} + ib(x,t) \tag{39}$$

into the propagator equation (28),

$$\Pi(\eta,\varepsilon;x,t) = (2\pi i D\varepsilon)^{-1/2} \exp - i\varepsilon b(x,t) \exp - \frac{1}{2} \varepsilon \frac{\partial u(x,t)}{\partial x} \exp \frac{i(\eta - u(x,t)\varepsilon)^2}{2D\varepsilon} \tag{40}$$

for small $\varepsilon$. The differential equation (30) and its complex conjugate (31) become

$$\frac{\partial \psi}{\partial t} = \frac{iD}{2} \frac{\partial^2 \psi}{\partial x^2} + u(x,t) \frac{\partial \psi}{\partial x} + \frac{1}{2} \frac{\partial u(x,t)}{\partial x} \psi - ib(x,t)\psi, \tag{41}$$

$$\frac{\partial \psi^*}{\partial t} = -\frac{iD}{2} \frac{\partial^2 \psi^*}{\partial x^2} + u(x,t) \frac{\partial \psi^*}{\partial x} + \frac{1}{2} \frac{\partial u(x,t)}{\partial x} \psi^* + ib(x,t)\psi^*. \tag{42}$$

It is noted that equations (4) and (40) are precise counterparts, as each represents a Gaussian propagator for which the norm of the propagated function is conserved: the former conserves the norm in **R**, and the latter conserves the norm in the sense of **C**.

Feynman [3] did not discuss how to maintain constancy of the norm to first order in $\varepsilon$ (norm conservation over time), but did not need to, because in his approach the necessary information is carried in from observational evidence: that is, in the Lagrangian, which gives the scalar and vector potentials, and in the midpoint rule, shown after alternative rules are unsuccessfully tried, to ensure the integrations match the known physics. We found the correct formulation purely by normalizing the Gaussian-propagated system, without reference to observational physics.

## 2.4 Checking alternatives

We have constructed the normalization constant when the Gaussian propagator equation (6) was restricted, by a "convenient" choice, to have real $D$ and $u$, with $D$ constant. We now go back over the equations critical for normalization to see how they stand when those restrictions are relaxed in any way possible.

Let us look at the terms in equation (33) with the stated variables being complex instead of real. Thus $D$ is to be replaced by $\mathrm{Re}D + i\mathrm{Im}D$, and $D^*$ by $\mathrm{Re}D - i\mathrm{Im}D$. Also $u$ is replaced by $\mathrm{Re}u + i\mathrm{Im}u$, and $u^*$ by $\mathrm{Re}u - i\mathrm{Im}u$; while $T$ and $T^*$ remain as before. Then if we collect the terms comprising real components of these variables, the integrations come to zero, as they did before, for the real components only; but the corresponding terms with imaginary components would fail to cancel because of altered signs in the complex conjugate terms. Also if $D$ is not constant with respect to $x$, $D$ cannot be taken outside the integral sign for the integration over $x$. If any one or more of these modifications were to be made, residual terms would be left, involving $\psi$, its derivatives, their complex conjugates, and imaginary components of parameters, unable to be simplified as a general case and therefore not identically zero for all square-integrable $\psi$. In such cases, the norm would not be conserved. Could this be rectified by setting $T$ to cancel the unwanted terms? No, because that would make $T$ a function of $\psi$, which would mean that propagation is not by a Gaussian, therefore not admissible for the present discussion.

As for $D$, it must be constant in space, but there is nothing in our equations that says it must be constant in time. That is the only relaxation in the restrictions that we have reason to identify. We will comment further on this in **2.5**.

Since our aim is to consider the Gaussian in a *fully* complex form, we must also consider how the independent variables (the space and time coordinates) are to be regarded when they too are written as complex numbers. Explicitly, each space and time coordinate would be represented by a complex plane instead of just an axis of real numbers. Usually we are only interested in $\Pi$ and $\psi$ over the real axes, so for most purposes the complex Gaussian is adequately represented with the space and time coordinates written as real, whilst keeping in mind that they could be treated as complex should that be needed.

## 2.5 Schrödinger equation

A few notational changes will align the equations with standard usage. We substitute

$$D = 1/m, \tag{43}$$

which is equivalent to $D = \hbar/m$ implying Planck's constant $\hbar = 1$. Also our use of *u* has implied that it is the *x* component of the vector **u**; we now denote this component as $u_x$ with

$$u_x = A_x D = \frac{A_x}{m}. \tag{44}$$

This equation deals with *x* components; we may write similar equations for *u* and **A** components in the *y* and *z* directions in considering three dimensions. Finally we replace $b$ by

$$b = \frac{\mathbf{A}^2}{2m} + \phi, \tag{45}$$

where $\mathbf{A}^2 = A_x^{\ 2} + A_y^{\ 2} + A_z^{\ 2}$.

These substitutions make the Schrödinger equation a little more complicated, but are aligned with common usage, and it will turn out that the Hamiltonian is made simpler when written in terms of $\phi$ instead of $b$. The substitutions imply no new conditions and no change in substantive meaning, because $A_x$ is defined at equation (44) in terms of $u_x$ and $D$, which are given as input parameters from the beginning, at equation (6) ($u_x$ is shown there as $u$); analogous input parameters $u_y$ and $u_z$ would define $A_y$ and $A_z$. The latitude enjoyed by $b$ is now carried by $\phi$.

Incorporating these changes into equations (41, 42),

$$\frac{\partial\psi}{\partial t} = \frac{i}{2m}\frac{\partial^2\psi}{\partial x^2} + \frac{A_x}{m}\frac{\partial\psi}{\partial x} + \frac{1}{2m}\left(\frac{\partial A_x}{\partial x}\right)\psi - \frac{iA_x^{\,2}}{2m}\psi - i\phi\psi. \tag{46}$$

We may write similar equations for components in the $y$ and $z$ directions, leading to

$$\frac{\partial\psi}{\partial t} = \left(\frac{i}{2m}\nabla^2 + \frac{1}{m}\mathbf{A}\cdot\nabla + \frac{1}{2m}\nabla\cdot\mathbf{A} - \frac{i}{2m}\mathbf{A}^2 - i\phi\right)\psi, \tag{47}$$

which is familiar as the Schrödinger equation [11] with $\phi$ and $\mathbf{A}$ representing scalar and vector potentials, as may characterise an electromagnetic field. Formal generalisation of the Schrödinger equation to three dimensions using the path integral derivation has been shown [12].

This is equivalent to the operator equation for the Hamiltonian,

$$H = \frac{1}{2m}(\mathbf{p} - \mathbf{A})^2 + \phi. \tag{48}$$

We remark in passing that the classical Gaussian, equation (4), permits asymmetry in that the steps may tend predominantly in a particular direction, as if steps are decided by a coin-toss using a coin with bias; net tendency is expressed by the variable $u(x,t)$ in equation (4) which thus describes a flow or drift. With the same equation written as equation (6) for the complex case, $u(x,t)$ again describes a net tendency or bias, but due to the algebra of complex functions the bias is manifested in Schrödinger's equation not as simple flow, but as the vector potential $\mathbf{A}(\mathbf{r},t)$, which is related to $u(x,t)$ through equation (44).

So we see that the form of the Schrödinger equation equation (46) is consistent with a process determined by a complex Gaussian propagator equation (6), with norm conservation.

We noted earlier that $D$, and hence $m$, could vary in time without violating norm conservation. This would however go against the conservation of mass, so the latter must be regarded as a separate law, not explained within the present non-relativistic arguments. We follow common practice in writing $m$, not $m(t)$, acknowledging that this implies that mass is constant in time, but is not here proven to be so. This remark also applies to previous derivations.

# 3. Discussion

## 3.1 Complex numbers and comparison with random walk

It has been long known that algebraic solutions for higher-order equations (cubic and quartic) involve square roots of quantities that are inevitably negative for certain ranges of values of the coefficients [13]. If we refused to deal with such "impossible" cases, there would be an artificial restriction on which situations we could consider, even when the solutions are in fact real and innocuous.

For this and similar reasons [2], it may be said that **C** has a more complete logic than **R** (as briefly suggested in the Introduction), which is of course borne out by the many applications of the complex number system.

In a Royal Society issue on the theme "Second quantum revolution: foundational questions", Cassinelli and Lahti [14] outlined an approach for an axiomatic reconstruction of quantum mechanics. Given that the basic structures of quantum mechanics are equally valid in each of the three cases of an infinite-dimensional Hilbert space over the real numbers, the complex numbers, or the quaternions, Cassinelli and Lahti argue that the real and quaternionic options both imply unnecessary complications when compared with the complex theory, and moreover that quaternionic quantum mechanics suffers from being unable to describe compound systems. The conclusion is that quantum mechanics is to be formulated in a complex Hilbert space.

It is known that the central limit theorem is a property of infinite dimensional separable complex Hilbert spaces [15, 16, 17]. This is of particular relevance here because it supplies a rationale for our choosing to investigate complex Gaussian propagators in the first place (equation 6 *et seq*.).

Concerning the assumption of norm conservation, we remark that a fundamental law of motion may indeed be required to conserve the norm, since otherwise the system would either collapse or explode in short order.

When irregular motion is calculated using ordinary numbers **R**, the random-walk equation (equation 4) is the known result [10]. As a new result, we have made the comparable calculation in complex numbers **C,** leading to Schrödinger's equation.

(It may avoid possible misunderstanding to note that a random walk in **C** is sometimes described as though it has two degrees of freedom in the Argand diagram, with steps in the real direction being independent of steps in the imaginary direction. That picture does not respect the one-dimensional character of a complex number, so it is not appropriate for our discussion.)

## 3.2 Born postulate

The continuous density functions, whether real or complex, progress deterministically in time. Use of the term "random" then needs to be justified. This is clear in the real case (the classic random walk), given that the equations have been deduced from random variable theory [10]. But the axioms of random variable theory are defined so as to be inapplicable to complex values, so the statistical interpretation is not so immediate. However, in the complex case, having arrived at the Schrödinger equation, we find that $|\psi|^2$ does after all have a statistical interpretation. Originally postulated by Born to account for quantum mechanical observations [11], this interpretation was later presented as a theorem within axiomatic approaches [19-22], tending to remove the Born rule as a separate postulate of quantum mechanics. Further, it has been argued as the only consistent way to interpret complex amplitudes [23, 24].

In a computer study of a dynamic origin for the Born probability rule in the de Broglie-Bohm pilot-wave theory, it was considered that the Born distribution should not be regarded as an axiom because it can be reproduced dynamically [25]. Thus, using Schrödinger's equation to calculate the dynamics, the Born quantum probabilities were efficiently approached, suggesting a status similar to thermal probabilities in ordinary statistical mechanics. The fact that Schrödinger's equation is here traced back to the central limit theorem suggests that such a picture includes the Schrödinger equation itself, i.e. the equation is not simply deterministic as generally taught.

## 3.3 Compare with previous justifications of Schrödinger's equation

Previous approaches to Schrödinger's equation have started with considerable information taken from observational physics.

Thus Schrödinger [5] drew on the conservation of energy, the de Broglie relations and Hamilton's analogy between waves and particles to construct wavefunctions. Feynman [3] used the Lagrangian to express action and develop path integrals. In both arguments a very large amount of physical knowledge is encompassed in these inputs.

In contrast we did not use any physical information as input: the physical content here appears entirely as output, after purely algebraic examination of the complex Gaussian propagator.

There is also a distinction with the use of complex numbers. Both Schrödinger [5] and Feynman [3, 4] introduced complex phase factors in the wavefunctions and path integrals arbitrarily, giving no *a priori* justification for doing so. We overcome that arbitrariness by allowing all the variables to be complex, and then recovering the Schrödinger equation by taking norm conservation into account. It can be claimed that the holistic use of **C** is not arbitrary, but can be justified by its more symmetrical and complete logic, and by critical consideration of the division algebras (see **3.1**).

Further to the primary sources [3-5], we single out three classic contributions of different kinds [6-8].

Kac [6] recognized that the diffusion equation is related to the Schrödinger equation by analytic continuation in the time variable. However he did not say why complex numbers should be used at all -- apart from empirical success -- or why time should be the variable distinguished in this way.

Nelson [7] attempted to reformulate quantum mechanics in terms of real statistical processes. Nelson was aware that an unsatisfactory feature of the model was its predication on continuous trajectories for the particles. Nelson also assumed Newtonian mechanics as given. Neither of those features is assumed here.

In a group-theoretic approach, Jauch [8] put forward as a theorem of unitary operators that Galilean kinematics constrains the Hamiltonian to take the standard form (48). A difficulty with this has been pointed out more recently. Brown and Holland [26] showed that it depends on a seemingly innocuous but nontrivial assumption at one particular group-theoretic step. They maintain that unless a satisfactory *a priori* justification is provided for this step, the foundation of the Jauch theorem is obscure;

other derivations made along the same lines are open to the same question ([26] and references therein).

The present derivation (Section **2**) is distinct from that of Jauch in that it does not make use of group theory, and does not assume Galilean kinematics; hence it is not open to the Brown and Holland criticism [26]. Rather the system is treated in one frame of reference only, finding the Schrödinger equation and Hamiltonian for that frame. For another frame of reference, one would repeat the calculations using coordinates for that frame, and find the comparable results. The derivation requires no transformation of coordinates between different reference frames.

More generally, regarding the complex Hilbert space formalism, Mackey [27] indicated that it rests on postulates that have never had prior justification: its use was "arbitrary … based on the practical consideration that it was known to work".

In summary, reliance on prior physical assumptions in some form, and the arbitrary and partial use of **C**, are consistent characteristics of previous approaches to the Schrödinger equation ([28] and references therein), which are avoided in the present approach.

### 3.3.1 Modern reconstructions

Because the axioms of the quantum mechanical formalism appear physically obscure and difficult to interpret, there were early reconstructions of quantum theory based on different sets of axioms. Many of these are referenced in Hardy [29], but they do not completely relieve the obscurities. Hardy reopened the debate [30] with the aim of finding axioms that have more direct physical or informational meaning. Examples are those of Hardy [30, 31], Rovelli [32], Clifton *et al*. [33], Rohrlich [34], Chiribella [35], Goyal [36], Spekkens [37].

The change from interpreting quantum mechanics to reconstructions of it has been described by Grinbaum [38] as a paradigm shift.

All the reconstructions rest on assumptions. Arguably the present assumptions are few in number, of a general nature and relatively simple: they consist of a Gaussian propagator, and norm conservation in **C**.

Randomness has been seen as a puzzle to which the many worlds interpretation [39] may provide a resolution. As the present treatment deals with Gaussian functions, which occur by virtue of the central limit theorem, randomness is implied, but it is indifferent as to possible sources of randomness.

# 4. Conclusion

It is concluded that the content of Schrödinger's equation is equivalent to propagation by a generalized Gaussian function, normalized in the sense of **C** with the norm conserved in time. The key role of the Gaussian may be attributed to the central limit theorem, which extends to random vectors in infinite-dimensional separable complex Hilbert spaces (**3.1**).

No physical laws need to be postulated *a priori* (such as the Lagrangian or other Newtonian constructs including the action formula; Galilean kinematics; or the de Broglie relations or equivalent). Since we did not need to start with physical laws, this supports the proposition that Schrödinger's equation is only a consequence of the properties of pure numbers (though we worked fully in the complex number system to elicit this).

A standard teaching is that Schrödinger's equation describes "deterministic" progress of the wavefunction, in contrast to "collapse" according to the Born probabilities. However the present connection to the central limit theorem suggests that Schrödinger's equation and the Born probabilities both belong to a picture that resembles statistical mechanics.

# References

1. Serway, R. A., Moses, C. J., Moyer, C. A.: *Modern physics* (Saunders College Publishing, 1989).
2. Stewart, I., Tall, D.: *Complex analysis: the hitchhiker's guide to the plane* (Cambridge University Press, 1983).
3. Feynman, R. P.: Space-time approach to non-relativistic quantum mechanics. *Rev. Mod. Phys*. **20**, 367-387 (1948).
4. Feynman, R. P., Hibbs, A. R.: *Quantum mechanics and path integrals* (McGraw-Hill, 1965).
5. Schrödinger, E.: Quantisation as a problem of proper values (Part IV) in *Collected papers on wave mechanics*. (ed. Schrödinger, E.) 102-123 (Blackie and Son, English translation, 1928).
6. Kac, M.: On distributions of certain Wiener functionals. *Trans. Amer. Math. Soc*. **65**, 1-13 (1949).
7. Nelson, E.: Derivation of the Schrödinger equation from Newtonian mechanics. *Phys. Rev*. **150**, 1079-1085 (1966).
8. Jauch, J. M.: *Foundations of Quantum Mechanics* (Addison-Wesley, 1968).
9. Van Kampen, N. G.: *Stochastic Processes in Physics and Chemistry* (North-Holland, 2007).
10. Gillespie, D. T.: *Markov processes: an introduction for physical scientists* (Academic Press, 1992).
11. Schiff, L. I.: *Quantum mechanics* (McGraw-Hill,1968).
12. Schulman, L. S.: *Techniques and applications of path integration* (Dover Publications, 2005).
13. Stillwell, J.: *Mathematics and its history* (Springer, 3rd ed. 2010).
14. Cassinelli, G., Lahti, P.: Quantum mechanics: why complex Hilbert space? in theme issue, "Second quantum revolution: foundational questions" (eds. Jaeger,G., Khrennikov, A., Perinotti, P). *Phil. Trans. R. Soc. A* **375**, 20160393 (2017).
15. Dedecker. J., Merlèvede, F.: The conditional central limit theorem in Hilbert spaces. *Stochastic Processes and their Applications*. **108**, 229-262 (2003).
16. Hoffman-Jorgensen, J., Pisier, G.: The law of large numbers and the central limit theorem in Banach spaces. *Ann. Probab.* **4**, 587–599 (1976).
17. Jain, N.C.: Central limit theorem and related questions in Banach spaces. *Proc. symposium in pure mathematics, American mathematical society* **XXXI**, 55–65 (1977).
18. Gardiner, C. W.: *Handbook of stochastic methods for physics, chemistry and the natural sciences* (Springer, 2003).
19. Gleason, A. M.: Measures on the closed subspaces of a Hilbert space. *J. Math. Mech.* **6**, 885-93 (1957).
20. Finkelstein, D.: The logic of quantum physics. *Trans. N.Y. Acad. Sci*. **25**, 621-37 (1962).
21. Hartle, J. B.: Quantum mechanics of individual systems. *Am. J. Phys*. **36**, 704-12 (1968).
22. Graham, N.: The measurement of relative frequency in *The many-worlds interpretation of quantum mechanics* (eds. DeWitt, B. S. & Graham, N.) 229-253 (Princeton University Press, 1973).
23. Caticha, A.: Consistency, amplitudes and probabilities in quantum theory. *Phys. Rev. A.* **57**, 1572-82 (1998).
24. Tikochinsky, Y.: Feynman rules for probability amplitudes. *Int. J. Theor. Phys*. **27**, 543-549 (1998).
25. Valentini, A., Westman, H.: Dynamical Origin of Quantum Probabilities. *Proc. R. Soc. A*. **461**, 253–272 (2005). doi:10.1098/rspa.2004.1394.

26. Brown, H. R., Holland, P. R.: The Galilean covariance of quantum mechanics in the case of external fields. *Am. J. Phys*. **67**, 204-214 (1999).
27. Mackey, G. W.: *Unitary Group Representations in Physics, Probability and Number Theory* 194 (Addison-Wesley, 1989).
28. Skorobogatov, G. A., Svertilov, S. I.: Quantum mechanics can be formulated as a non-Markovian stochastic process. *Phys. Rev. A*. **58**, 3426-3432 (1998).
29. Hardy, L.: Reconstructing quantum theory. arXiv:1303.1538v1 [quant-ph] (2013).
30. Hardy, L.: Quantum theory from five reasonable axioms. arXiv:quant-ph/0101012v4 (2001).
31. Hardy, L.: Reformulating and reconstructing quantum theory. arXiv:1104.2066v3 [quant-ph] (2011).
32. Rovelli, C.: Relational quantum mechanics. *International Journal of Theoretical Physics*. **35**, 1637–1678 (1996).
33. Clifton, R., Bub, J., Halvorson, H.: Characterizing quantum theory in terms of information theoretic constraints. *Foundations of Physics* **33**, 1561-1591 (2003).
34. Rohrlich, D.: Three attempts at two axioms for quantum mechanics. arXiv:1011.5322v3 [quant-ph] (2010)
35. Chiribella, G., D'Ariano, G. M., Perinotti, P.: Informational derivation of quantum theory. arXiv:1011.6451v3 [quant-ph] (2011).
36. Goyal, P.: Information Physics—Towards a New Conception of Physical Reality. *Information* **3**, 567-594 (2012).
37. Spekkens, R.: In defense of the epistemic view of quantum states: a toy theory. Manuscript at arXiv:quant-ph/0401052v2 (2005).
38. Grinbaum, A.: Reconstruction of quantum theory. *The British Journal for the Philosophy of Science.* **58,** 387-408 (2007).
39. Vaidman, L.: Quantum theory and determinism. *Quantum Studies: Mathematics and Foundations*. 1, 5-38 (2014). https://doi.org/ 10.1007/s40509-014-0008-4